\documentclass[a4paper]{aa}
\usepackage{graphicx}
\usepackage{epsfig}
\usepackage{latexsym}
\usepackage{txfonts}
\usepackage{natbib}
\bibpunct{(}{)}{;}{a}{}{,}

\begin{document}
\renewcommand{\labelitemi}{-}
\title{Observations of conduction driven evaporation in the early rise phase of solar flares}
 \author{Marina Battaglia \inst{1} \and Lyndsay Fletcher \inst{2} \and Arnold O. Benz \inst{1} }
\institute{Institute of Astronomy, ETH Zurich, 8093 Zurich, Switzerland \and Department of Physics and Astronomy, University of Glasgow, Glasgow G12 8QQ, UK}
\date{Received /Accepted}

\abstract
{The classical flare picture features a beam of electrons, which were accelerated in a site in the corona, hitting the chromosphere. The electrons are stopped in the dense chromospheric plasma, emitting bremsstrahlung in hard X-rays. The ambient material is heated by the deposited energy and expands into the magnetic flare loops, a process termed chromospheric evaporation. In this view hard X-ray emission from the chromosphere is succeeded by soft-X-ray emission from the hot plasma in the flare loop, the soft X-ray emission being a direct consequence of the impact of the non-thermal particle beam. However, observations of events exist in which a pronounced increase in soft X-ray emission is observed minutes before the onset of the hard X-ray emission. Such pre-flare emission clearly contradicts the classical flare picture.}
{For the first time, the pre-flare phase of such solar flares is studied in detail. The aim is to understand the early rise phase of these events. We want to explain the time evolution of the observed emission by means of alternative energy transport mechanisms such  as heat conduction.}
{RHESSI events displaying pronounced pre-flare emission were analyzed in imaging and spectroscopy. The time evolution of images and full sun spectra was investigated and compared to the theoretical expectations from conduction driven chromospheric evaporation. }
{The pre-flare phase is characterized by purely thermal emission from a coronal source with increasing emission measure and density. After this earliest phase, a small non-thermal tail to higher energies appears in the spectra, becoming more and more pronounced. However, images still only display one X-ray source, implying that this non-thermal emission is coronal. 
The increase of emission measure and density indicates that material is added to the coronal region. The most plausible origin is evaporated material from the chromosphere. Energy provided by a heat flux is capable of driving chromospheric evaporation. We show that the often used classical Spitzer treatment of the conductive flux is not applicable. The conductive flux is saturated. During the preflare-phase, the temperature of the coronal source remains constant or increases. Continuous heating in the corona is necessary to explain this observation.  }
{The observations of the pre-flare phase of four solar flares are consistent with chromospheric evaporation driven by a saturated heat flux. Additionally, continuous heating in the corona is necessary to sustain the observed temperature. }

\titlerunning{Preflares}
\authorrunning{Marina Battaglia}

\maketitle

\section{Introduction}

The question of energy conversion during the impulsive phase of a solar flare has converged over the course of the past three decades on a picture featuring an acceleration site in the corona, where particles accelerated to high energies then precipitate along magnetic field lines to the chromosphere.
 In the chromosphere they lose their energy in the dense plasma, leading to chromospheric heating, as well as the characteristic hard X-ray `footpoint' emission. In addition to radiating and conducting
its excess energy away, the heated chromospheric plasma finds a new equilibrium by expanding up the loop, in a  process termed 'chromospheric evaporation'. The observational evidence for a relationship between flare heating and non-thermal electrons in the impulsive phase is reasonably strong. In this paper we use imaging and spectroscopy from the RHESSI satellite \citep{Li02}, as well as information from the GOES satellites to investigate flare heating before the flare impulsive phase, apparently in the absence of non-thermal electrons.

Chromospheric evaporation is usually proposed as the cause of flare extreme ultraviolet (EUV) and soft X-ray (SXR) emission \citep[though see ][who dispute this on several grounds]{Fe90,Ac92}. \\
The main observational arguments for chromospheric evaporation in flares are (1) that the density of the coronal plasma observed to be emitting in SXRs during flares is one or two orders of magnitude greater than is generally measured in the quiet corona \citep[eg.][]{AA01,Kr08}, requiring a source for the additional material and (2) the presence of upflowing plasma detected via blueshifts in high temperature spectrum lines \citep{An82}, in particularly those observations made with imaging spectrometers such as the Coronal Diagnostic Spectrometer \citep[CDS,][]{Ha95} on the Solar and Heliospheric Observatory \citep[SOHO,][]{Dom95}. Such flows have been observed in the flare impulsive phase by \citet{Mi06b,Mi06a} and in the gradual phase by \citet{Cza99}. Evaporative flows have been described as being either `gentle' or `explosive', with the latter occurring when the rate of chromospheric plasma heating greatly exceeds the rate at which it can cool by radiation, conduction or expansion, primarily determined by the ratio between the heating timescale and the hydrodynamic expansion timescale \citep{Fi85}. The relatively slow heating rate by conduction from a heated corona makes it a likely source of gentle evaporation.  \citet{An78} developed an analytic model of conductive driven evaporation in the decay phase of flares, though report that the `evaporation' represented the hydrodynamic redistribution of material already at coronal temperatures, rather than chromospheric material being heated and expanding into the corona. \citet{Kar87} carried out numerical simulations exploring the effects of non-local and saturated heat flux in addition to classical Spitzer heat flux.
Observational evidence for such ``gentle'' evaporation was found by \citet{Za88} in observations from the Solar Maximum Mission, and the flows observed by \citet{Mi06b} and \citet{Cza99,Cz01} were also found to be consistent with conductive evaporation, due to both their relatively low speeds and absence of significant HXR radiation.

In the electron beam-driven evaporation model, the thermal X-ray emission is a direct effect of the energy deposition in the chromosphere by the electron beam, so that the time-integrated non-thermal hard-X-ray (HXR) flux in a given energy range is proportional to the SXR flux. This was first proposed by \citet{Ne68} and since then, the so-called `Neupert effect' has been studied at length by various authors \citep[e.g.][]{De93,McT99,Ve05}. Overall the expected proportionality between the time-derivative of the SXRs and the instantaneous HXRs is supported observationally, particularly for impulsive flares, though there is a significant scatter interpreted as due to energy losses from low-temperature radiation, conduction and mass motion. There are also some significant exceptions from this rule -  for example, \citet{McT99} find that the Neupert relationship manifested in about half of the 33 flares they studied. A similar result was found by \citet{Ve02ap}.  This suggests additional energy input, not related to non-thermal electrons, in the non-Neupert flares.  Various authors have also observed impulsive SXR footpoint emission tracking the HXR intensity, rather than its integral \citep[e.g.][]{Hu94}. Finally, pre-flare SXR sources, occurring in advance of any HXR emission present a further instance in which thermal emission may also be completely unrelated to non-thermal electrons. This paper concerns the observation and interpretation of such sources.

It has been known for some time that flare-related activity commences prior to the flare impulsive phase. There is a distinction to be drawn between pre-flare activity, which refers to the very earliest stages of the flare before the impulsive phase radiation is detectable, and `flare precursor' events, which are small-scale brightenings in UV to HXR wavelengths happening some tens of minutes before the flare. The first to use the term ``pre-flare'' and complete a statistical study on flare precursors in X-rays were \citet{Bu59}, who inferred the X-ray behavior via ionospheric disturbances. With the arrival of the \emph{Yohkoh} satellite \citep{Og91}, the nature of the pre-flare phase was studied in more detail, particularly the relationship between the non-thermal and thermal emissions.
It was noted relatively early that high temperature thermal sources were present in the corona substantially before the impulsive phase flare onset \citep{Ac92}, a result
which is clearly at odds with the assumption that electron beams drive evaporation.
Comprehensive statistics including an analysis of the spatial relations between pre-flares and the consequent flares has been conducted by \citet{Fa96} and \citet{Fa98} based on observations with the Yohkoh Soft X-ray Telescope \citep[SXT,][]{Ts91}. Their definition of pre-flare activity is a rise in the \emph{Yohkoh}  emission above background,  five minutes to an hour before the main peak. In the spatially-resolved  SXT images it is clear that in several of their sample there is substantial coronal soft X-ray emission occurring several minutes before the start of the impulsive phase \citep{Fa98}. In a study of 10 \emph{Yohkoh} flares, \citet{Al98} also identify line broadening observed with \emph{Yohkoh} Bragg Crystal Spectrometer \citep{Cu91} several minutes before the onset of the impulsive phase, and \citet{Ha01} show that in one such event the temperature, intensity, and non-thermal line width are changing significantly, well before the start of the hard X-rays. Taken together, the \emph{Yohkoh} observations provide strong evidence for pre-flare coronal activity unrelated to evaporation driven by electron beams. In recent years, RHESSI observations further confirmed the existence and relative relative frequency of pre-flare activity as shown eg. in \citet{Ve02}.

The Ramaty High Energy Solar Spectroscopic Imager \citep[RHESSI,][]{Li02} offers new opportunities to study pre-flare sources in detail. RHESSI is a HXR telescope, with spectral coverage extending down in energy to the thermal range at $\sim$ 3-20~keV. Its high spectral resolution (of about 1~keV at energies up to 100 keV) permits spectroscopic diagnostics not available previously, coupled with high spatial resolution (as low as 2.3", depending on the flare) in user-defined energy channels. This offers new temperature and emission measure diagnostics for very hot plasmas (around 10 MK and above) as well as for non-thermal electrons, if present. The RHESSI observations presented here display an increase of low energy X-ray emission, which can be fitted with a thermal spectrum, up to minutes before the rise of the HXR emission.

The morphology of these events shows only one source, visible at energies in the thermal regime, while the footpoints only appear at the onset of the HXR emission. We will argue that this indicates that another mechanism of energy transport to the chromosphere may be important, such as heat conduction. \\


In this study, we present the first comprehensive RHESSI-led study of the pre-flare phase of solar flares.  The paper is structured in the following way: Section \ref{selection} describes the flare selection process. In Sect.~\ref{tevol} the time evolution of the selected flares is analyzed in spectra and images. A theoretical model to explain the observed time evolution is presented in Sect.~\ref{theory}, followed by a Discussion and Conclusions.

\section{Flare selection} \label{selection}
\begin{figure*}
\includegraphics[height=10cm,width=16cm]{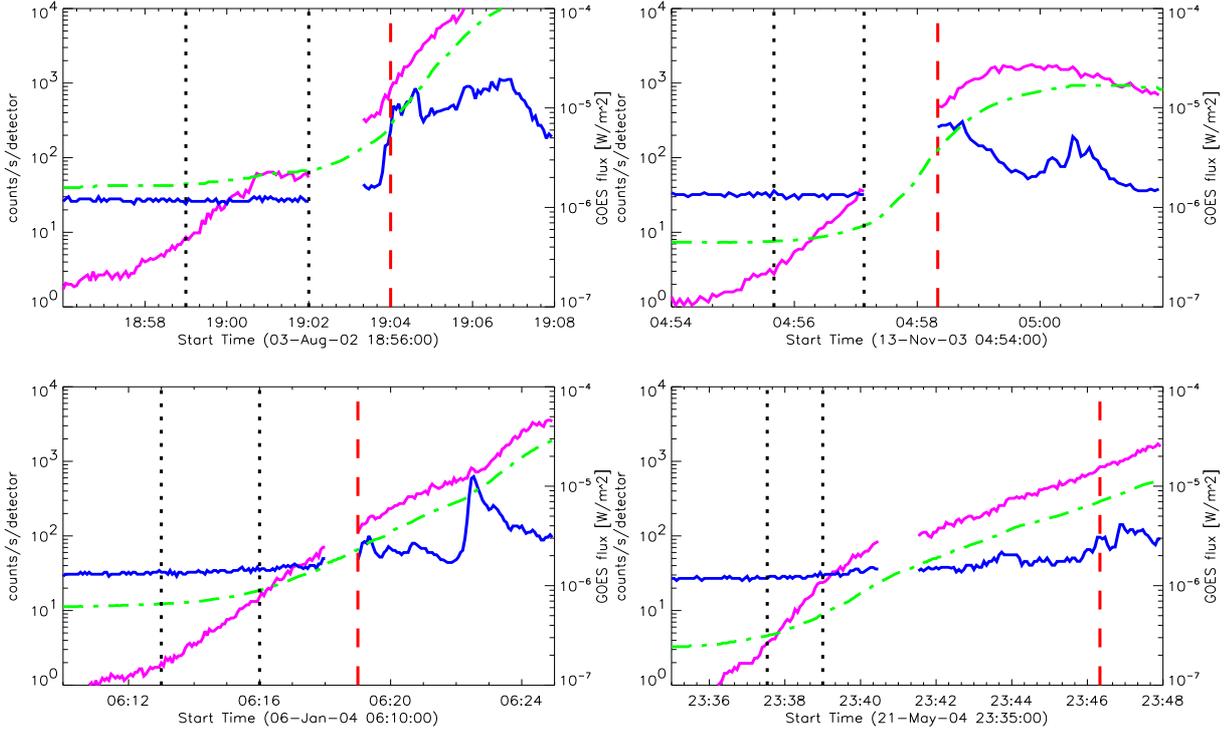}
\caption {RHESSI corrected count rate lightcurves in 6-12 keV and 25-50 keV during the pre-flare and early impulsive phase of the events. Missing data in the RHESSI lightcurves indicate the time interval that was omitted in the analysis due to attenuator state change. The \textit{dotted} vertical lines indicate the time interval of the primary analysis, the \textit{dashed} line gives the start time of the image in which footpoints are observed first. GOES lightcurve (\textit{dash-dotted}). }
\label{lcurves}
\end{figure*} 

In this study, events with pre-flare activity will be analyzed. The main pre-requisite of such a flare is increasing emission in SXR well before the onset of the HXR emission. Therefore, only events the beginning of which was fully observed were considered. The selection criteria are summarized here:
\begin{itemize}
\item Offset of more than 700$"$ from disk center to limit projection effects in the study of the flare morphology;
\item GOES class larger than M1 to ensure good count rates for imaging;
\item The beginning of the event had to be observed with RHESSI;
\item SXR \& GOES emission had to start increasing at least 1 min before the HXR emission;
\item Events with a simple morphology involving one SXR source at the onset and two HXR sources appearing later in the event. This indicates the time when the acceleration process becomes dominant;  
\item High enough detector livetime ($>$ 90 \%) for reliable spectroscopy and imaging. 
\end{itemize}

Four events were selected as best suited for this kind of study. Their key parameters are listed in Table~\ref{events}. 
The RHESSI attenuators were out during the earliest phase of all events with the thin attenuator moving in as the count rates increased, typically after 2-3 minutes. 
 \begin{table*}
\caption{Eventlist. The attenuator state is 0 for all events at the very beginning, then changes to 1 as the flux increases. The attenuator times give the rough time interval that must be avoided because of attenuator motion. }
\begin{center}
\begin{tabular}{lllllll}

Date & Analyzed time& GOES class &xy-position [arcsec]& coronal source volume [cm$^3$]&Attenuator&Attenuator times \\
\hline
03-Aug-2002 & 18:59-19:02&X1.2& 912/-271&4.9$\cdot 10^{26}$&0/1& 19:02:20/19:03:10 \\
13-Nov-2003& 04:55:40-04:57:12 &M1.7& -977/23&4.8$\cdot 10^{26}$&0/1&04:57:20/04:58:20\\
06-Jan-2004&06:13-06:16 &M5.9& -992/88&2.8$\cdot 10^{27}$ &0/1&06:18/06:19\\
21-May-2004&23:37:30-23:39 &M2.6& -757/-157&7.5$\cdot 10^{26}$&0/1&23:40:30/23:41:30\\
\hline

\end{tabular}\end{center}

\label{events}
\end{table*}

\section{Time evolution of events} \label{tevol}
The time evolution of the SXR emission in GOES and RHESSI 6-12 keV along with the HXR emission (RHESSI 25-50 keV) is shown in Fig.~\ref{lcurves}. The missing data in the RHESSI lightcurves indicate the time intervals omitted due to attenuator state change. The attenuator state was 0 for all events at the onset of the flare.\\ 

\subsection{Data analysis}
We studied the time evolution of the pre-flare phase both in images and spectra. CLEAN and Pixon images \citep{Hur02,Me96} during the pre-flare and early impulsive phase were made, using grids 3 to 8. The time interval of each image was 30s in order to get high enough count rates throughout the pre-flare phase. 
Full-sun photon spectra were fitted in 30 s time intervals, corresponding to the image time intervals. The fitting model consisted of a thermal component fitted from 6 keV and, if possible, a non-thermal component. Spectra from the onset of the flares as well as the start of the non-thermal emission are shown in Fig.~\ref{spectra}. The RHESSI spectrum at the lowest energies (3-6 keV) could be fitted with an additional thermal component with temperature and emission measure similar to the values that can be derived from GOES measurements (not shown in the figure).  \\

\subsection{Spatial evolution} \label{spatialevol}
\begin{figure*}
\centering
\includegraphics[width=17cm]{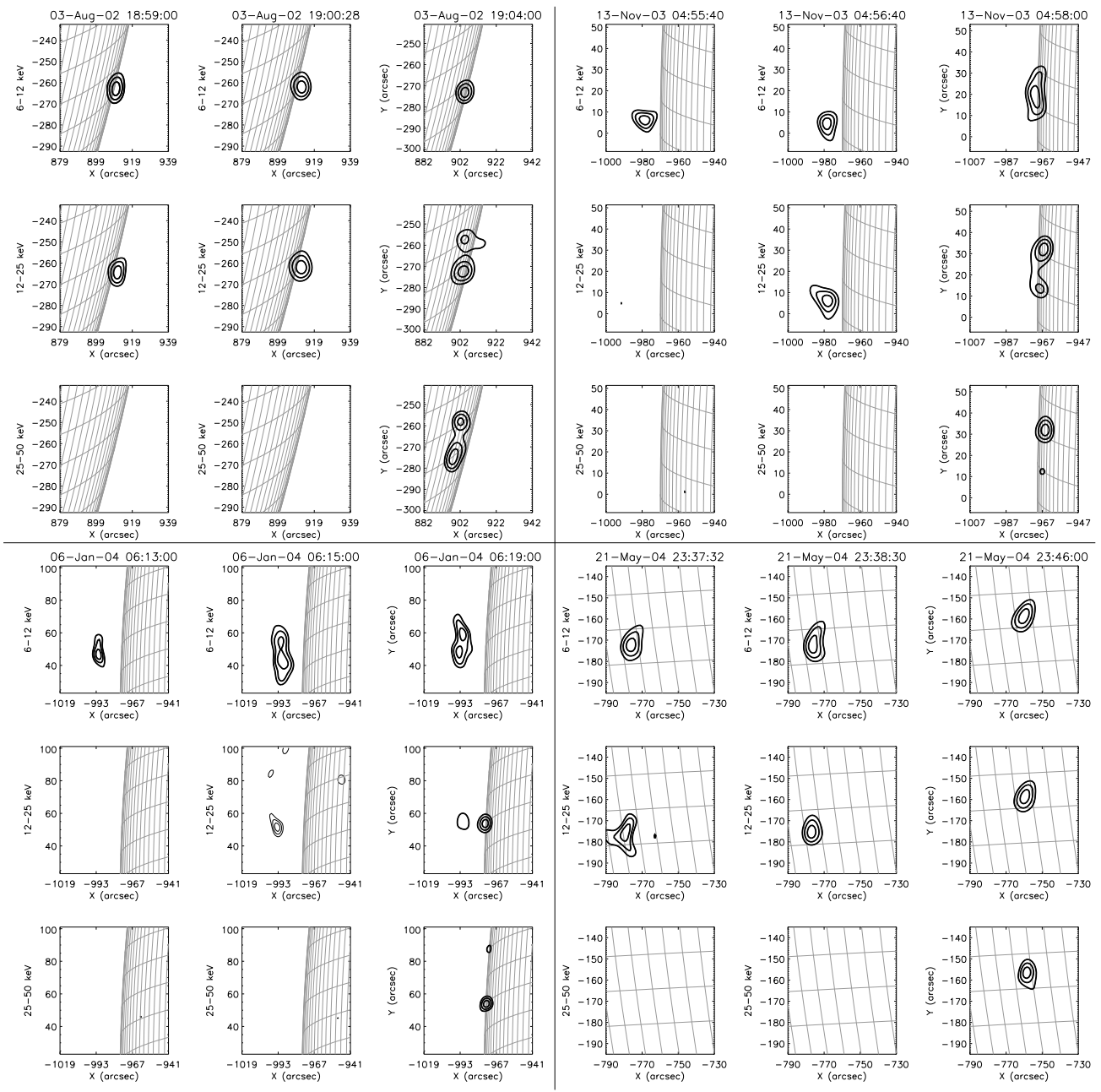}
\caption{Contour plots from CLEAN images (60\%, 75\% and 90\% of maximum flare emission) of all events taken at the first fit interval of the rise phase (\textit{left}), the time interval when a non-thermal component could first be fitted (\textit{middle}) and at the first appearance of the footpoints (\textit{right}). Those times correspond to the times of the spectra shown in Fig.~\ref{spectra}. The energy bands are 6-12 keV (\textit{top}), 12-25 keV (\textit{middle}) and 25-50 keV (\textit{bottom}).}
\label{cleanimg} 
\end{figure*}

\begin{figure}
\resizebox{\hsize}{!}{\includegraphics{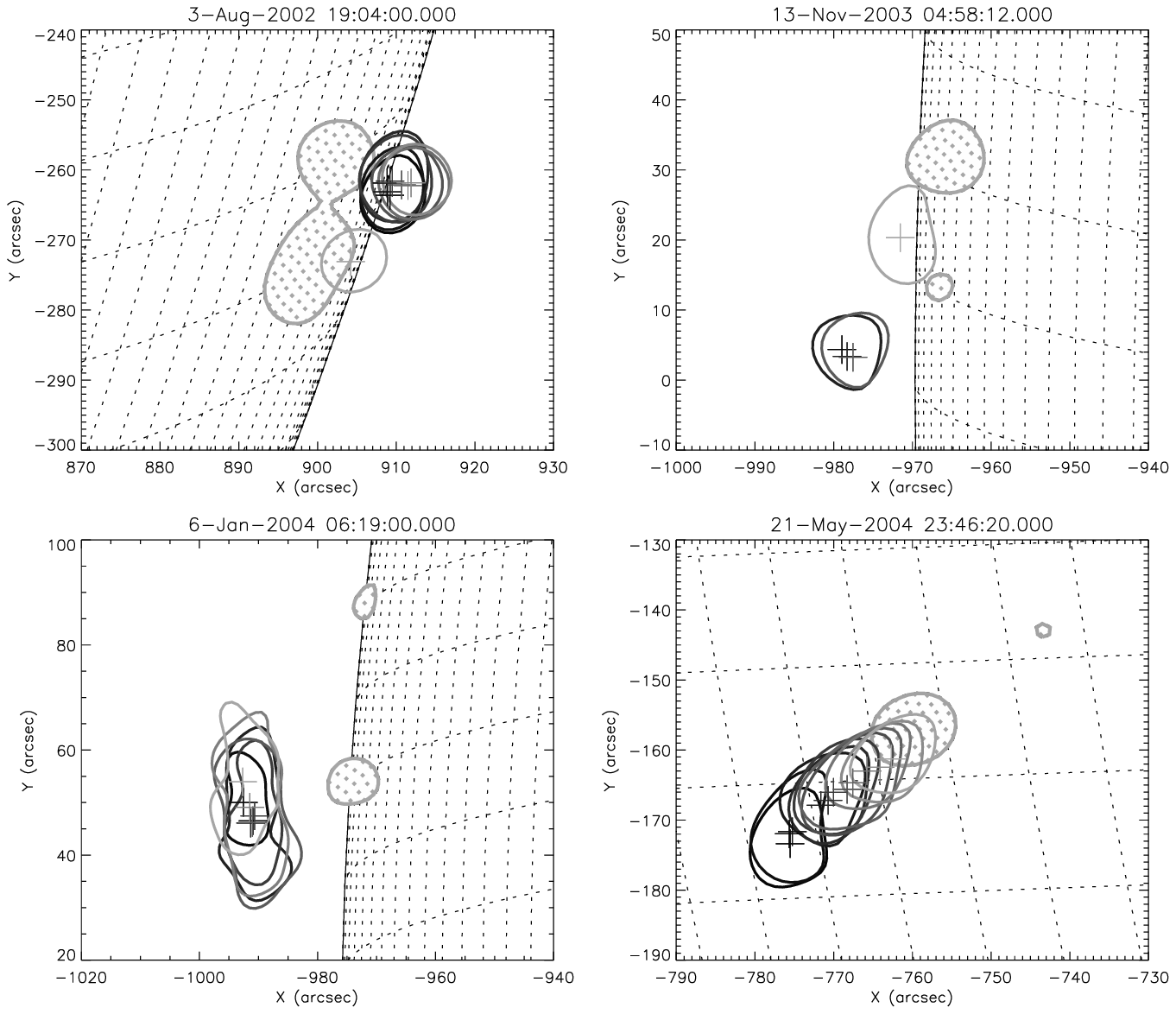}}
\caption {Position of the coronal source in time in the energy range 6-12 keV. Time evolution is indicated by 70\% intensity contours in colors ranging from black to light with the first appearance of the footpoints (25-50 keV) given as dotted areas in the appropriate color. The centroid position is marked with a \textit{cross} in the corresponding color.}
\label{sourcemotion}
\end{figure}
 Figure~\ref{cleanimg} shows CLEAN images of the events taken at the beginning, at the time when a non-thermal component was first fitted and at the time when the footpoints first appear. For each event, 3 energy-bands (6-12, 12-25, 25-50 keV) are shown. 

All events start with a single source visible only at the lowest energies in the range 6-12 keV (upper left image of the 9-image panel per flare in Fig.~\ref{cleanimg}. This is interpreted as a source at the top of a loop (coronal source). After some minutes two additional sources appear at higher energies (25-~50 keV) which are interpreted as chromospheric footpoints (lower right image of each flare in Fig.~\ref{cleanimg}). In three events the position of the first appearing source is clearly displaced from the footpoint position, implying that there is an actual loop geometry with a coronal source on top and footpoints in the chromosphere. In the fourth event, the separation is not as clear but from the spectral evolution (see next Section) one can still assume a coronal-source-footpoints-geometry.

\subsubsection{Movement of coronal sources}

We analyzed the time evolution of the source position by measuring the centroid position of the 50\% contour in CLEAN and Pixon images in the energy range 6-12 keV. The position from both imaging methods agree within the uncertainties. Figure~\ref{sourcemotion} displays the 70\% contour from CLEAN images with positions as found from CLEAN. Colors go from black to light (time of first appearance of footpoints). 
The positions shown indicate the position of the coronal source from the beginning until the time of the first appearance of footpoints, in the same time steps as used in spectroscopy (30s). 

In all events, the sources move to some extent during the pre-flare phase. The source in the event of 03-Aug-2002 is stable within the position uncertainties during the pre-flare phase, but appears at a displaced location in the last image (at the first appearance of the footpoints). Due to the attenuator state change just before the last image, an interval of about 1 minute is missing. Therefore, one cannot say whether the source moves continuously or not. The same observation is made in the 13-Nov-2003 event. Here, the sources observed during the pre-flare phase and at the beginning of the impulsive phase could possibly indicate two separate loop systems as suggested by \citet{Liu06}.  The source position  of the 6-Jan-2004 event remains constant over time. The 21-May-2004 event is somewhat peculiar, displaying pronounced, continuous source motion. The displacement of the position is clearly larger than the uncertainties. From the direction of the movement, this could be another case of altitude decrease as has been found in other RHESSI events \citep[eg.][]{Sui03,Ve06}

\subsection{Spectral evolution} \label{specevol}

\begin{figure*}
\centering
\includegraphics[height=10cm, width=17cm]{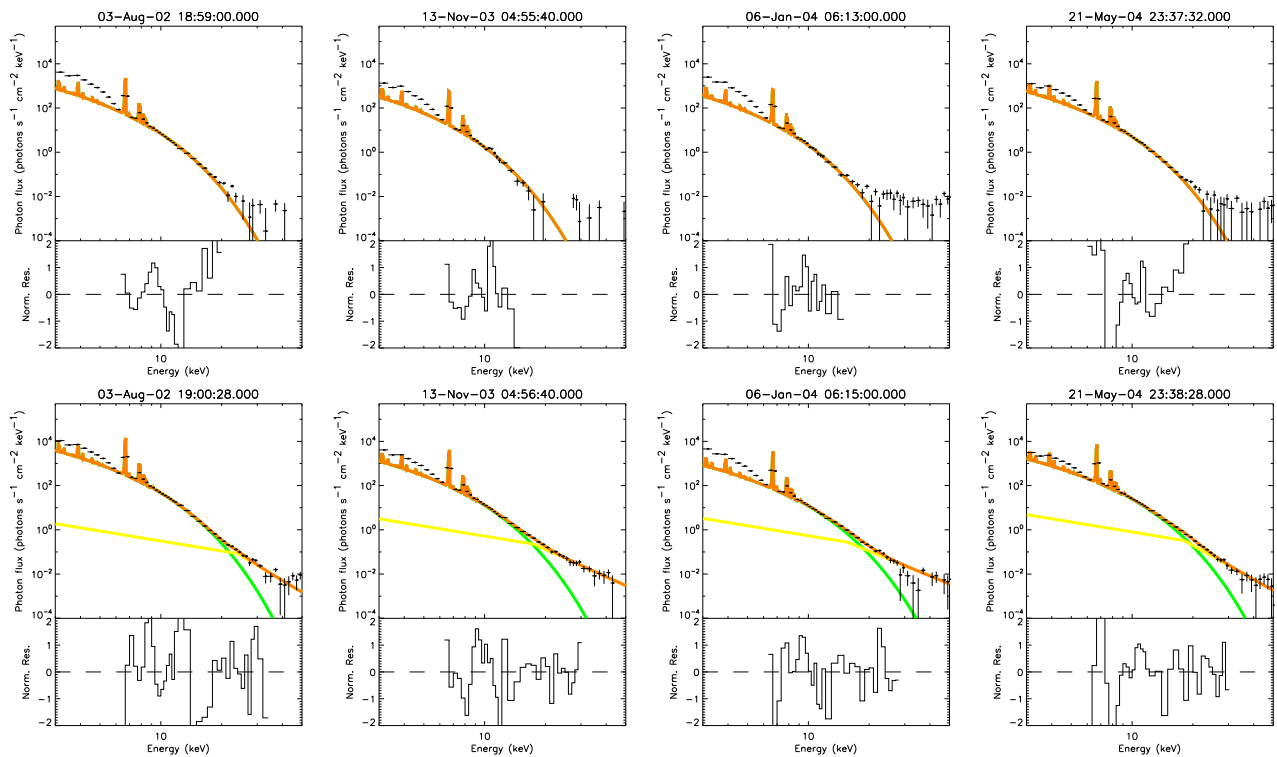}
\caption {Spectra with model fits for 2 time intervals. \textit{Top}: First analyzed time interval. \textit{Bottom}: First interval during which a non-thermal component could be fitted. }
\label{spectra}
\end{figure*}

All four events display a purely thermal spectrum in the first of the analyzed time intervals. This spectroscopic finding is further supported by GOES SXI images (not shown here). These are available for three out of four events and show a soft X-ray source minutes to hours before the RHESSI observations. In RHESSI observations the apparently purely thermal phase lasts one to two minutes, after which a small tail to higher energies becomes visible (Fig. \ref{spectra}). At this stage, there are as yet no footpoints are observed, indicating that the non-thermal emission is coronal. The tail appears independent of albedo correction and as the events are all near the limb, the influence of albedo on the total flux is minimal. Further, pile-up is not large enough at this stage to account for the emission. It is therefore safe to assume that the tail is real. It can be fitted with a power-law or with a second, very hot thermal component. During the impulsive phase (after attenuator state change) the HXR component becomes very pronounced and clearly distinguishable. Figure~\ref{spectra} shows spectra of the selected events, one taken at flare onset and one when the non-thermal tail first appears. 

\begin{figure*}
\centering
\includegraphics[height=16cm]{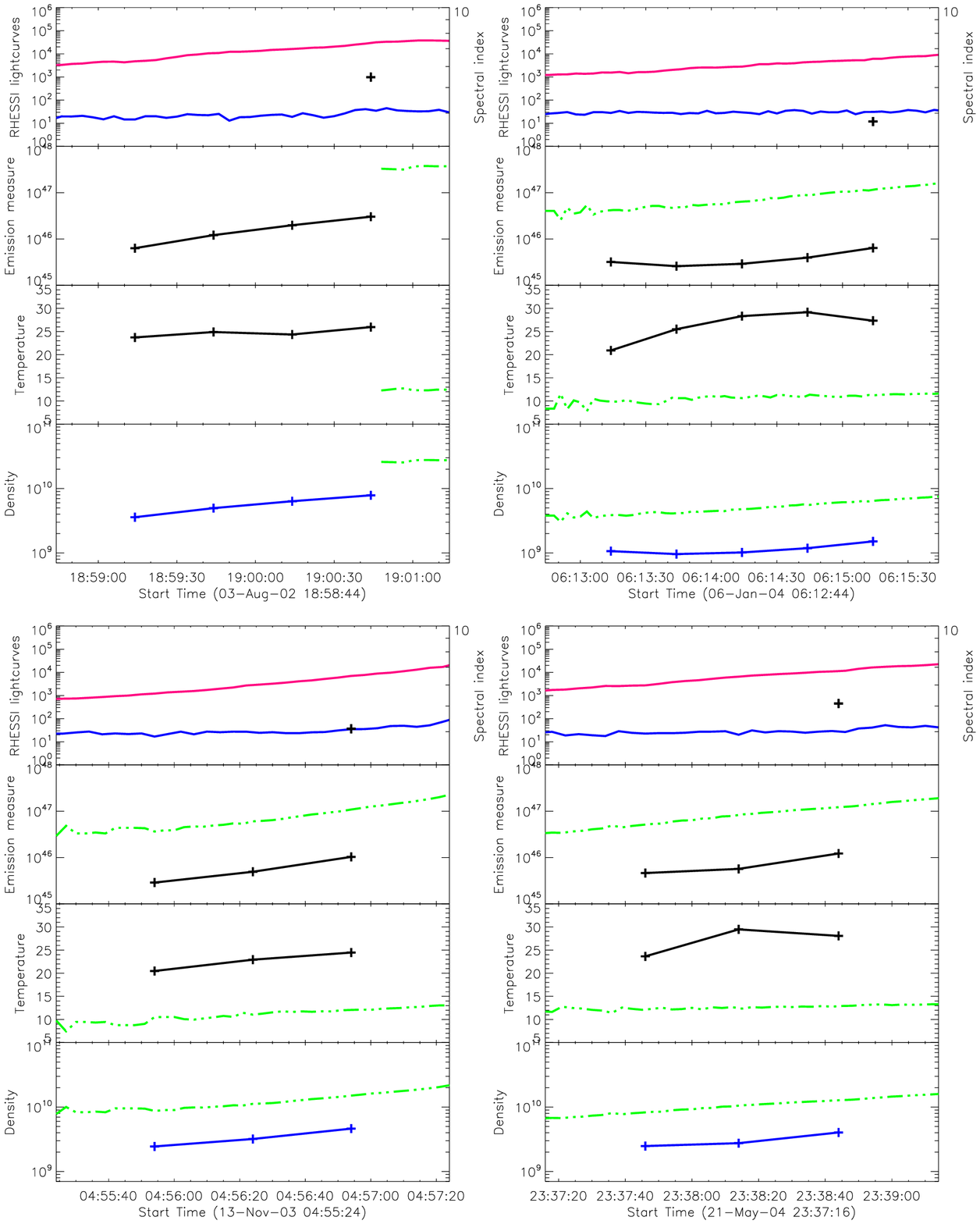}
\caption {Time evolution of the fit parameters for all 4 events. \textit{Top}: RHESSI lightcurves in the 6-12 and 25-50 keV energy band. The cross indicates first appearance and hardness of the non-thermal component. \textit{Second}: Emission measure as computed from RHESSI (\textit{crosses} and GOES \textit{dash-dotted}). \textit{Third}: Temperatures, same as for emission measures. \textit{Fourth}: Electron density in the coronal source, same as for emission measures.}
\label{fitparams}
\end{figure*}

\subsection{Time evolution of flare parameters}
The time evolution of the fit parameters, namely temperature, emission measure and spectral index is shown in Fig.~\ref{fitparams}. The figure displays panels for each event. The top panel of each event shows RHESSI lightcurves in the 6-12 and 25-50 keV energy band. The cross indicates the time of first appearance, and the hardness, of the non-thermal component. The second panel displays the emission measures as determined from RHESSI spectra and GOES. In the third panel, temperatures are presented the same way as the emission measures. The fourth panel shows the electron density estimates. The critical parameter in computing densities is the flare volume. RHESSI images allow an estimate of the volume by measuring the source area $A$ and approximating the volume as $V=A^{3/2}$. Measuring the area from RHESSI images is quite imprecise with large uncertainties. We determined the 50 \% contours in CLEAN images (de-convolving the CLEAN beam). The area is constant over the pre-flare evolution and we made an estimate of the quantitative value of the volume by taking the time average of the measured values. The average electron density in the coronal source is then given as
\begin{equation}
n_e\approx\sqrt{EM/V}\,,
\end{equation}
assuming a volume filling factor of 1.
\section{A theoretical model to explain the observations} \label{theory}
In Fig.~\ref{fitparams} we present the time evolution of flare parameters such as temperature, emission measure and density, as well as images. The observations reveal increasing emission measure and densities over the course of the pre-flare as well as generally increasing temperatures. Assuming a constant,  or even increasing, source volume, the density increase can only be attributed to  additional material which is added to the coronal source region. The most likely scenario causing such a density increase is chromospheric evaporation. In this scenario, chromospheric plasma (including the transition region) is heated and expands upward the magnetic loops. There are two possible main heating mechanisms for the chromosphere; non-thermal electron beams and thermal conduction. In the former case, a beam of high energetic electrons impinges on the chromosphere, where the particles are completely stopped in the dense target. As a consequence of this energy deposition, the temperature rises. In the latter case, energy is transported from the hot coronal source to the cooler chromosphere by thermal conduction. We are going to discuss the thermal conduction scenario further in the following subsection. Reasons why thermal conduction is favored over particle beams are presented in the discussion. 

\subsection{Theory of thermal conduction}\label{condtheory}

Our model will explain chromospheric evaporation by heating due to energy input from heat conduction. The coronal source is much hotter than the chromospheric plasma, therefore a temperature gradient develops, driving a heat flux downward along the magnetic loop. 

The conductive heat flux is given as 
\begin{equation}
F_{cond}=\kappa_0T^{5/2}\frac{\partial T}{\partial s}.
\end{equation}
In classical thermal conduction theory following \citet{Spbook}, the conductive coefficient is $\kappa_0=10^{-6}$ erg cm$^{-1}$s$^{-1}$K$^{-7/2}$. For a flare of half loop length $L_{loop}$ and coronal source temperature $T_{cs}$, this may be approximated by
\begin{equation}
F_{cond}\approx 10^{-6}\frac{T_{cs}^{7/2}}{L_{loop}}\quad\quad \mathrm{[erg\, cm^{-2} s^{-1}]}.
\end{equation}

\begin{figure}
\centering
\resizebox{\hsize}{!}{\includegraphics{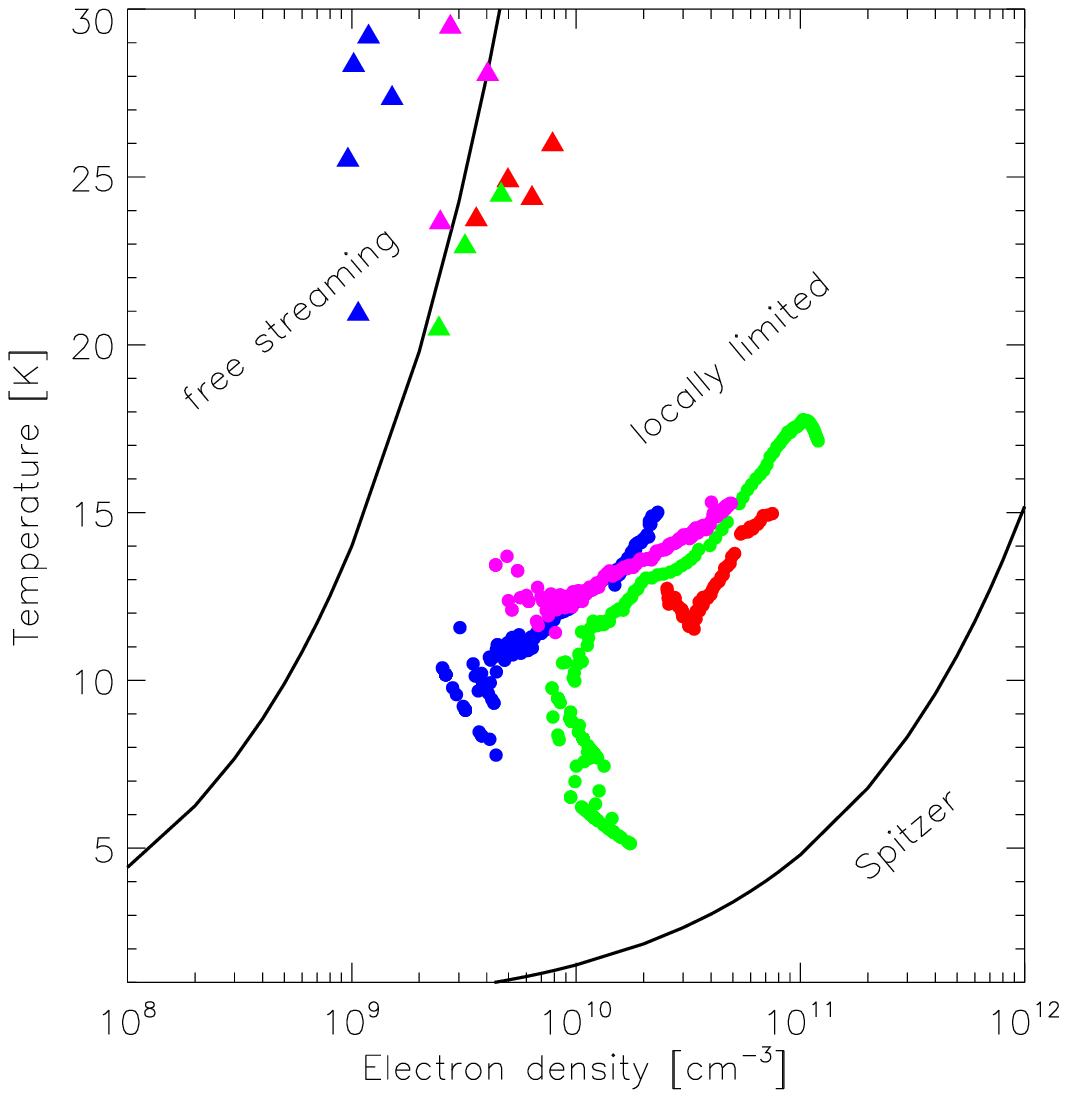}}
\caption {Temperature vs. density space of typical flare loop values, indicating the boundaries and regimes for classical thermal conduction, flux limited and non-local conduction. The density and temperature values of the observed events are given as \textit{dots} for GOES and \textit{triangles} for RHESSI observations. Different shades stand for the different events.}
\label{noncllocal}
\end{figure}

However, the maximum heat flux a plasma can carry is limited by a fraction of the thermal energy-flux $F_{therm}=n_ekT_ev_{th}$, where $v_{th}$ is the thermal electron velocity. For sufficiently large temperature gradients,  such as occur in solar flares, the heat flux is expected to reach this limit and saturate.
\citet{Gra80} showed that there are two regimes of flux saturation. A non-classical treatment with local flux limiting is necessary if the electron mean free path $\lambda_{emf}=5.21\cdot 10^3 T^2/n_e$ \citep{Bebook} is larger than only 0.12\% of the temperature scale length $L_{th}$. If the electron mean free path even exceeds the temperature scale length, the heat flux becomes non-local in the sense that it depends on the global density and temperature structure of the plasma. Figure~\ref{noncllocal} illustrates the boundary conditions for locally limited and non-local heat flux in the density-temperature plane of typical solar flare values.
The locally limited regime is described by \citet{Ca84} as a continuous transition from the purely classical treatment to the non-local regime. Depending on the ratio  $\mathcal{R}=\lambda_{emf}/L_{th}$, a reduction factor $\varrho<1$ is applied to the classical heat flux, resulting in an effective heat flux of $F_{red}=\varrho (\mathcal{R})\cdot F_{cond}$. \citet{Ca84} published the values of $\varrho$ for different $\mathcal{R}$ and atomic number Z. Using the values for Z=1, we fitted a function of the form
\begin{equation}
\varrho=A\cdot e^{-b(x+c)^2}
\label{corrfact}
\end{equation}
to the values published by \citet{Ca84}. The parameters are $x=\ln (\mathcal{R})$, A=1.01, b=0.05, c=6.63. Using those values, the locally limited, reduced heat flux can be computed for any $\mathcal{R}$ from Eq.~(\ref{corrfact}). 

In the non-local regime where the electrons are freely streaming and complete saturation is reached, the flux can be expressed as:
\begin{equation}
F_{sat}=0.53n_em_ev_{th}^3\quad \mathrm{erg\, cm^{-2} s^{-1}}
\end{equation}
\citep{Kar87}.

\subsection{Observation of thermal conduction} \label{obsthcond}
\begin{figure}
\centering
\resizebox{\hsize}{!}{\includegraphics{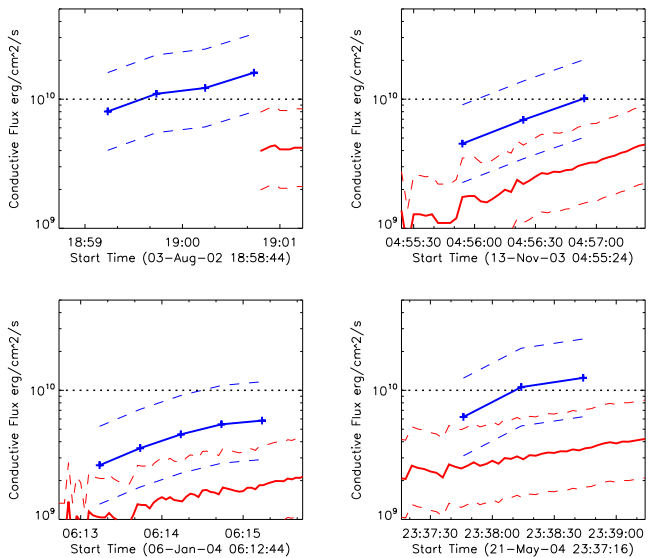}}
\caption {Effective conductive flux derived from RHESSI (\textit{crosses on solid line}) and GOES (\textit{solid line}). The \textit{dottet} line marks the threshold energy flux to trigger explosive evaporation \citep[from][]{Fi85}. The \textit{thin dashed} lines mark the level of uncertainty (see \ref{cond:discussion}).}
\label{condfluxes}
\end{figure}

In studies of thermal conduction in flare loops, a classical scenario involving Spitzer conduction is often assumed. Are the conditions in the observed pre-flares consistent with classical conduction or is the flux limited? The densities and temperatures as derived from GOES and RHESSI observations are displayed in Fig.~\ref{noncllocal}. Values measured with GOES are given as dots, RHESSI measurements are marked as triangles. GOES typically yields larger emission measures and smaller temperatures than RHESSI. Assuming the same emission volume, this results in higher densities for GOES. The figure reveals that the conditions during the observed times of the pre-flare cannot sustain a classical heat flux, the heat flux is limited. The resulting effective flux was computed from the observed values. Temperatures and densities yield the electron mean free path $\lambda_{mfp}$. The temperature scale length $L_{th}$ is given as $L_{th}=T/\Delta T \approx L_{loop}$, where $\Delta T$ was approximated as $\Delta T \approx T_{cs}/L_{loop}$. Using an estimated loop half-length of $10^9$ cm we computed the ratio $\mathcal{R}$. The correction factor and therefore the effective heat flux was computed from Eq.~(\ref{corrfact}). The results are displayed in Fig.~\ref{condfluxes}, again for both, GOES and RHESSI values.  

\subsection{Chromospheric evaporation}
We assume that the heat flux computed in the section above is deposited in the transition region and chromosphere, heating the chromospheric material. The hot material will then expand upward  the magnetic loops. 
This chromospheric evaporation is  usually divided into two types; gentle evaporation and explosive evaporation. The determining factors distinguishing the two are the ratios between heating, radiative cooling and expansion rate in the chromosphere. If the heating rate is much smaller than the radiative loss rate, all the deposited energy is radiated away. If the heating rate is only marginally larger than the radiative loss rate, the temperature and pressure rise slowly and the material evaporates gently. If the deposited energy rate exceeds a certain threshold value, the chromospheric plasma is heated to coronal temperatures quickly, a large overpressure develops and explosive expansion of plasma is the consequence. \citet{Fi85} derived a value for the critical heating flux of 10$^{10}$ erg cm$^{-2}$s$^{-1}$ for typical coronal and chromospheric conditions. It is generally assumed that explosive evaporation is associated with the large energy input provided by beams of fast electrons hitting the chromosphere, while gentle evaporation is normally attributed to heating by thermal conduction.  

\subsubsection{Conduction driven chromospheric evaporation}
Can the observed emission measure and density increase in the presented events be explained with chromospheric evaporation driven by the heat flux presented in Sect~\ref{obsthcond}? The total energy flux in the evaporated plasma is given as
\begin{equation}
F_{up}^{evap}=3n_e^{evap}kT^{evap}v_{up}=3m_en_e^{evap}v_{th}^2v_{up} \quad \mathrm{[erg\, cm^{-2} s^{-1}]},
\label{fup}
\end{equation}
where $v_{up}$ is the velocity of the upward moving plasma, $v_{th}$ the thermal velocity and $n_e^{evap}$ the density of the upflowing plasma. 

We use the  evolution of the density in the coronal source to estimate $v_{up}$. 
\begin{eqnarray} 
\frac{\mathrm{d}n_e}{\mathrm{d}t}&=&\frac{\partial n_e}{\partial s}\frac{\partial s}{\partial t} \approx  \frac{n_e^{evap}}{L_{loop}}v_{up} \\
 v_{up} &\approx & \frac {L_{loop}}{n_e^{evap}}\cdot \frac{\mathrm{d}n_e}{\mathrm{d}t}  \label{upflowvel} 
\end{eqnarray}
The density of the upflowing plasma cannot be determined from observations. Assuming $n_e^{evap}=n_e$, the density in the coronal source, the upflow velocities derived from Eq.~(\ref{upflowvel}) range from 20 km s$^{-1}$ to 150 km s$^{-1}$. This results in an evaporative flux of the order of 10$^8$ erg $\mathrm{cm^{-2} s^{-1}}$. Further, the evaporative expansion requires an energy of 
\begin{equation}
E_{exp}=\int pa \mathrm{dl}
\end{equation}
where p is the total pressure. Assuming isothermal expansion the expansion energy flux can be computed \citep[see][]{Kb02}:
\begin{equation}
F_{exp}=2m_en_ev_{th}v_{up}\ln(L_{loop}/l_i)
\end{equation}
The expansion energy flux is larger than the thermal energy flux (Eq. \ref{fup}) by a factor of $2/3ln(L_{loop}/l_i)$ which amounts to about 5. 

The total energy flux for evaporation is therefore in the range of $5\cdot 10^{8}$ erg $\mathrm{cm^{-2} s^{-1}}$, which is well below the values obtained from the data (Fig.~\ref{condfluxes}). The rest of the conducted energy heats the transition region and upper chromosphere to sub-coronal temperatures and is radiated away.  

\section{Discussion} \label{cond:discussion}
\subsection{Electron beams in the pre-flare phase}
 In the above section, we demonstrated that conduction driven chromospheric evaporation can account for observations of pre-flares. 
Can electron beams below the detection limit do the same? Let us estimate the maximum expected non-thermal electron flux and the expected chromospheric heating rate.
 We estimate the electron beam strength starting from the observed coronal source spectrum. The first time-interval when non-thermal emission from the corona was observed was used as an estimate of the upper limit to the electron flux, assuming a thin target in the coronal source. If a beam constituting of this electron flux hits the chromosphere, a thick target footpoint spectrum $F_{fp}=AE^{-\gamma}$ is expected. From observations of footpoints and coronal sources \citep[eg.][]{Ba06}, we know that a flux $F_{fp}$ as found from this estimate would be well observed in images and spectra which is not the case. Taking the observations of \citet{Ba06} as a reference, we estimate that the effective electron flux that reaches the footpoints has to be about a factor 100 smaller than expected from the coronal source observations. For the computation of the beam heating rate, we therefore use an assumed footpoint photon flux spectrum $F_{fp}^{eff}=0.01\cdot F_{fp}$.

The heating rate of a beam of electrons was derived by \citet{Br73a} and \citet{Li76}. The rate of energy input per unit volume at column density $N$ and time $t$ is given as 
\begin{eqnarray}
I_B(N,t)&=&10^7n(N,t)[x(N,t)+0.55] \nonumber\\ 
&\times&\frac{A}{a}C(\gamma)(1.1\cdot10^{-17}N)^{-(\gamma+1)/2} \quad\quad \mathrm{[erg\,cm^{-3}]}
\label{nonthenrate}
\end{eqnarray}
where $n(N,t)$ is the hydrogen density, $x(N,t)$ the fractional ionization, $a$ the cross-section area, $A$ the normalization of the observed photon spectrum, $\gamma$ the spectral index of the observed photon spectrum and $C(\gamma)$ a function of the $\beta$-function and $\gamma$. The parameters of the expected footpoint spectrum were used as input for $A$ and $\gamma$ in Eq.~(\ref{nonthenrate}). As an estimate of the column density, the stopping depth of a 35 keV electron was used according to $N=10^{17}E_{\mathrm{kev}}^2$ \citep{Ta88}. The resulting density $n(N,t)$ and fractional ionization $x(N,t)$ was taken from \citet[][Table1]{Br73a}. The cross-section area was taken as $10^{17}$ cm$^2$, which corresponds to typical footpoint areas as measured by RHESSI. Finally, the factor $C(\gamma)$ was taken from Fig. 2 in \citet{Li76}. Inserting the above values gives a total energy input of the order of 3-4 erg cm$^{-3}$s$^{-1}$. From the expression of the column density $N=nl$, where $l$ is the path length, l can be determined and by multiplying Eq.~(\ref{nonthenrate}) with $l$ we get an energy input of 5-7$\cdot$10$^7$ erg cm$^{-2}$s$^{-1}$, which is two orders of magnitude smaller than the energy input by thermal conduction. Even if the effective footpoint area was an order of magnitude smaller, as often found in UV or H$\alpha$ observations \citep[eg.][]{Te07}, the heating rate by the beam would still be an order of magnitude smaller than the energy input by thermal conduction. Therefore, if there is any chromospheric heating by electron beams in the pre-flare phase, its contribution is minimal compared to the energy input by thermal conduction.  
\subsection{Influence of assumptions}
In all computations throughout this work, the values measured in the coronal source were used also for the loop legs. For some parameters, e.g. the electron mean free path $\lambda_{mfp}$, those values may not be accurate. For a complete treatment, the temperature and density in the loop is needed. However, those values cannot be determined observationally. If the loop was at the same temperature and density as the coronal source, it would be visible in X-ray images. As no loop emission is observed in the images, the actual loop has to be either less dense or cooler than the coronal source. Thus the coronal source values represent an upper limit to the true conditions in the loop. Smaller temperature could result in an order of magnitude smaller $\mathcal{R}$. On the other hand, a smaller density could cause an order of magnitude larger ratio $\mathcal{R}$ of electron mean free path to temperature scale length. The correction factor $\varrho$ (Eq. \ref{corrfact}) would change by 0.5 to 2, respectively and so would the conductive flux. The resulting upper and lower limits of the conductive flux are indicated by dashed lines in Fig.~\ref{condfluxes}. Even if the conductive flux differed by a factor of 2, this would not change the basic interpretation.

The evaporated flux $F_{up}^{evap}\sim n_eTv_{up}\sim T$ only depends on the temperature. It is therefore expected to be at most about a factor of 4 smaller. However, the upflow velocities derived from Eq.~(\ref{upflowvel}) are proportional to $1/n_e$ and could therefore be up to an order of magnitude larger for smaller densities. Eq. (\ref{upflowvel}) may then give velocities higher than the local sound speed which would be unphysical. 

\subsection{Mode of conduction}
Our results indicate that the conductive heat flux during the pre-flare phase is non-classical. The temperature gradient between the coronal source region and the chromosphere is so steep that the classical Spitzer conduction is not valid. The conductive flux is locally limited for temperature and density values determined from GOES and may be fully saturated (i.e. the electrons are freely streaming) in the case of the RHESSI measurements. Both situations lead to a reduced heat flux compared to the classical Spitzer conductivity. This finding does not change directly the conclusions drawn from the work presented here. However, it requires that care be taken when computing total flare energy budgets as the energy loss of the coronal source due to conduction may be smaller than anticipated from the assumption of classical Spitzer conductivity. 

\subsection{Acceleration vs. heating}
Despite the energy loss in the corona due to the heat flux, increasing temperatures during the pre-flare phase are observed. Continuous heating in the coronal source is therefore necessary to sustain and even increase the observed temperature. Possible heating mechanisms include heating at the stand-off slow-mode shocks in Petschek reconnection models which has been proposed as an explanation of the formation of hot thermal sources in Yohkoh flares \citep[eg.][]{Ts97}, or betatron heating in a collapsing magnetic trap \citep[eg.][]{Karl04}. A strong candidate heating mechanism is wave-particle interaction as described in the transit-time-damping model \citep{Mi96,Gr06}. Such a model could also explain the spectral time evolution of the observations: Depending on the escape time and wave density in the acceleration region, particles are not accelerated to non-thermal energies but remain at a quasi-Maxwellian distribution observationally not distinguishable from a purely thermal plasma at enhanced temperature. For increasing wave density or escape time, particles are accelerated to higher energies, constituting a non-thermal tail which is clearly noticeable in the observations. 

However, the three dimensional geometry of the flare site has to be considered and a smooth transition from heating to acceleration at the same site may not reflect reality in two of the presented events. As shown in Fig.~\ref{sourcemotion}, the position of the coronal source changes over time in the events of 03-Aug-2002 and 13-Nov-2003. \citet{Fa96} and \citet{Fa98} made a statistical study on the position of Yohkoh pre-flares and the successive impulsive flares, finding that only in 25\% of the observed events were the pre-flare emission and impulsive flare emission spatially coincident. The other events were classified into what the authors termed ``distant'' and ``adjacent/overlapping'' events. In the latter case, parts of the pre-flare 50\% maximum intensity emission overlap with parts of the main flare emission while in the ``distant'' events, the two emission patterns are spatially separated. If our two flares indeed fall into the category of distant events, it will not change the interpretation of the pre-flare phase, but care should be taken when interpreting the time-evolution of the spectrum beyond the purely thermal phase.   \\

\section{Conclusions}
The observed increase in thermal emission and SXR emission measure up to minutes before the start of the impulsive flare phase in the four events presented here can be explained by chromospheric evaporation driven by thermal conduction. To compensate for the heat loss, continuous heating in the coronal source is needed. In this early phase, acceleration of particles to non-thermal energies is minimal or non-existent. A scenario following stochastic acceleration is conceivable in which the coronal source region is only heated in the early flare stages. As soon as the acceleration mechanism becomes efficient enough, a tail of non-thermal particles is produced, first visible in the coronal source and eventually in the footpoints.

\mdseries
\begin{acknowledgements}
RHESSI data analysis at ETH Z\"urich is supported by ETH grant TH-1/04-02 and RHESSI software development by the Swiss National Science Foundation (grant 20-113556). This work was supported in part by Rolling Grant ST/F002637/1 from the UK's Science and Technology Facilities Council and by the European Commission through the SOLAIRE Network (MRTN-CT-2006-035484). This research made use of NASA's Astrophysics Data System Bibliographic Services and the RHESSI Experimental Data Center (HEDC). \end{acknowledgements}

\bibliographystyle{aa}
\bibliography{mybib}

\end{document}